\documentstyle[aps,twocolumn]{revtex}
\input{epsf}

\def\be{\begin{eqnarray}}
\def\ee{\end{eqnarray}}
\def\a{\alpha}
\def\b{\beta}
\def\d{\delta}
\def\D{\Delta}
\def\h{\chi}
\def\g{\gamma}

\def\f{\varphi}
\def\l{\lambda}
\def\o{\omega}

\def\bg{b^{\dagger}}
\def\cg{c^{\dagger}}

\def\dg{\dagger}
\def\ol{\overline}
\def\p{\partial}
\def\k{\kappa}
\def\s{\sigma}

\def\t{\theta}
\def\ur{\uparrow}
\def\dr{\downarrow}
\def\lrr{\leftrightarrow}
\def\bg{b^{\dagger}}
\def\cg{c^{\dagger}}
\def\e{\epsilon}

\begin{document}
\title{ Coexisting superconducting and spiral spin order: models of
ruthenate.}
\author{A.A.Ovchinnikov}
\address{ Max-Planck Institute of Physics of Complex systems,
N\"othnotzer Str., 38, D-01187 Dresden, Germany\\
Joint Institute of Chemical Physics of RAS,
Kosygin str.,4, 117334, Moscow, Russia\\}
%
\author{ M.Ya.Ovchinnikova}
\address{Joint Institute of Chemical Physics of RAS,
Kosygin str.,4, 117334, Moscow, Russia}
\date{03.05.2002}
\wideabs{
\maketitle
\begin{abstract}

Influence of a spiral spin structure on the superconducting (SC) pairing in
the most probable active $\g$ band of $Sr_2RuO_4$ is studied in the
mean-field approximation. Such structure with "incommensurate" vector
$Q=2\pi(1/3,1/3)$ has been proposed as that which removes the nesting
instability in $\a,\b$ bands. The pairing interaction of adjacent centers of
two
types - with attraction in singlet channel or in both the singlet and triplet
channels - is adopted.  In both cases the mixed singlet and triplet SC order
is revealed. The d-wave singlet pairing is accompanied by formation of the
p-wave triplet pairs $(k,-k-Q)_{\ur\ur}$ and $(k,-k+Q)_{\dr\dr}$ with large
total momenta $\mp Q$ and spin projections $\pm 1$ onto an axis normal to the
spin rotation plane of the spiral structure. Both the superconducting and
normal states have a broken time reversal symmetry which might be
detected by polarised photoemission.
\end{abstract}
}

\section{Introduction}
The problem of interplay between the superconducting and spin orders is
actual for all compound with strong electronic correlations. Among them a
single-layer quasi-2D ruthenate attracts considerable attention as a
superconductor
($T_c\sim 1.5K$)  with a possible triplet type of pairing \cite{1,2}.
One of the arguments in favor of such type of pairing is the Knight shift
behavior \cite{3}. It was suggested also that the pairing is governed by
the ferromagnet (FM) fluctuations which certainly exist in the ferromagnetic
parent compound $SrRuO_3$. The theoretical proposal \cite{5,5a}
of a spin-triplet p-wave superconducting (SC) order parameter (OP)
$\D_{ss'}(k)=(i\s_2\s_i)_{ss'}d_i(k)$ with $d_z(k)\sim k_x+ik_y$ is supported
by the observed temperature independent Knight shift \cite{3} and an increase
of muon spin relaxation below $T_c$ \cite{6}. This OP corresponds to the
nodeless gap function on the quasi-2D Fermi surface of $Sr_2RuO_4$. Similar
nodeless solution is a natural choice in a weak-coupling theory \cite{5a}.
However, the power temperature dependences as $T\to 0$, found for the
specific heat, $C(T)\sim T^2$ \cite{7}, NQR relaxation rate, $T_1^{-1}\sim
T^3$ \cite{8}, the thermal conductivity $\k  (T)\sim T^2$ \cite{9,10}, the
penetration depth \cite{4}, and ultrasonic attenuation \cite{11} point to
the node lines of the SC gap. In this connection the other possible gap
symmetries have been discussed \cite{12,13,14,15}. In particular, the
f-wave symmetry gap function with a horizontal node plane have been proposed
in \cite{12}. It seems to support the observed 4-fold symmetry of anisotropic
thermal conductivity of $Sr_2RuO_4$ under the in-plane magnetic field
\cite{9,10}.  But the observed anisotropy is consistent also
with a conventional d-wave pairing argued in some works.  However,
the latter hypothesis would require a new interpretation of the Knight shift
behaviour.

The situation with both the SC order and the magnetic properties might be much
more complicated if a normal state of the $RuO_4$ plane displays the spiral
(static or dynamic) spin structure. The latter has been proposed recently
\cite{16} in order to describe the incommensurate peak in a spin
susceptibility $\h ''(q,\o)$ at $q\sim Q=2\pi(1/3,1/3)$ observed  in
inelastic neutron scattering \cite{17,18} and some features in ARPES data
\cite{19} for $Sr_2RuO_4$. The properties of the quasi-1D sheets of Fermi
surfaces (FS's) for the $\a$ and $\b$ valence bands at total occupancy 4
electron per site   imply the nesting instability at $q=Q$ \cite{20,21}.
A spiral structure with $q=Q$ removes this instability for both quasi-1D
bands simultaneously even in zero order with respect to large on-centre
interaction. In mean field approximation the energy of spiral state with $Q=q$
is really lower than that of the para-, ferro-, antiferro-magnetic MF states
\cite{16}. The possibility of a coexistence of the SC pairing with a spiral
order, as well as of the coexistence of an antiferromagnet (AF) and SC orders
in cuprates, is yet an intriguing problem. It is interesting also since the
spiral spin structure means a time-reversal symmetry  breaking (TRSB) even
in the normal state and a rich set of new mixed states with the coexisting
singlet and triplet SC order parameters appear in the system.
A new perspective of
experimental study of TRSB phenomena has been open recently  by an
exciting results for the cuprates obtained with use of a photoemission (ARPES)
with polarised photons \cite{22}.

The aim of present work is to  study a possibility of coexistence of the
spiral spin order and superconductivity in models having relation with
$Sr_2RuO_4$. The symmetry and interplay of the singlet and triplet OP's are
studied. It is shown that both types of pairs survive simultaneously in
SC state in presence of spiral spin structure.  It is confirmed that the
most probable "active" band is the $\g$ band.  In this  band the preferential
singlet $d$-wave pairs coexists with the triplet pairs. Earlier
\cite{15} the possibility of mixing SC order in $Sr_2RuO_4$ has been supposed
on base of very close condensation energies for states with different
symmetries of pairing. It was considered that the pairing is mediated by spin
supceptibility peaked at incommensurate momentum. Unlike \cite{15} we derive
a microscopic mixing of the d-wave singlet and p-wave triplet SC orders
caused by the local spiral spin structure. The pairing is induced by model
interaction of adjacent centres. First we test the models with the
large pairing constants $\k$ and corresponding large transition temperatures.
Then we extend a study to more realistic models and calculate the
phase curve $T_c(\k)$ in full range of pairing constant.

Preliminary several basic points  of consideration must be
outlined.

1.Since the SC pairing interaction is supposed to be of electronic nature, we
try to model it by electronic interactions of adjacent sites of type
\be
V=\sum_{<nm>,\nu}V_{\nu}n_{\nu n}n_{\nu m}
   +J_{\nu}{\bf S}_{\nu n}{\bf S}_{\nu m}
\label{1}
\ee
for each of three bands $\nu=\a,\b, \g$. This corresponds to taking into
account the lowest $k$- harmonics in a momentum representation of pairing
interaction $V_{kk'}$, like it has been done in \cite{5}.
Interaction (1) corresponds to
the SC pairing constants $\k^s=2V+J/2$, $\k^t=2V-3J/2$ in the singlet
and triplet channels. The natural signs
$V>0$, $J>0$ would be expected from an experience in field of strongly
correlated systems. In the one-band models such signs of $V,J$ correspond to
$\k^t>0$, $\k^s<0$ in the triplet and singlet
channels and  thus they are more appropriate for a singlet, but not triplet
pairing. Here we extend a study on the both signs of pairing constant $\k^t$
in triplet channel.

2. The electronic structure of $Sr_2RuO_4$ is determined by three almost
independent $\a,\b,\g$ bands based on the d-orbits of $Ru^{+2}$ of the
$xz,~yz,~xy$ nature \cite{20,21}. Only small hybridisation of xz and yz bands
takes place at crossing points of their Fermi surfaces (FS's) at $k_x=\pm
k_y$.  According to \cite{5,5a} the orbital symmetry significantly
suppresses also the interband Cooper pair scattering which induces the SC gap
simultaneously in all sheets of FS's. So, a study is divided
into the following stages.  First one is a study of the SC pairing
and its symmetry in each band separately and a selection of a most probable
"active" band for SC instability. Then the interband scattering of Cooper
pairs and the proximity effect \cite{5} should be taken into account to
provide a superconductivity in the whole three-band system. Here we concern
only a first stage of problem.

3. In difference from \cite{5} we start from the normal MF state with a
broken time reversal symmetry, namely with a local spiral spin structure
characterised by a diagonal vector $Q=2\pi(1/3,1/3)$. This is a normal state
with the non-zero spin currents $j_{\ur}=-j_{\dr}$ of opposite directions for
two different spin polarisations perpendicular to the spin rotation plane of
the spiral state. This means that the electrons with up (down) polarisations
occupy preferentially the k-states with ${\bf kQ}<0$ or ${\bf kQ}>0$
correspondingly. Such symmetry leads to the polarisation asymmetry of FS's
revealed in \cite{16} and it can lead also to formation of the mixed singlet
and triplet pairing in SC state of system.
\section{MF treatment  in a spiral spin configuration}
A three-band model of $RuO_4$ plane is described by Hamiltonian \cite{21}
\be
H=T+H_U+V;~~T=\sum_{\nu,\s}\sum_k \e_{\nu k}\cg_{\nu k\s}c_{\nu k\s}
\label{21}
\ee
$$
H_U=\sum_{n,\nu } \Bigl\{ U n_{\nu n\ur}n_{\nu n\dr} +
\sum_{\nu '\neq\nu}[U_2{1\over 4}n_{\nu n}n_{\nu' n}
-J{\bf S}_{\nu n}{\bf S}_{\nu' n}] \Bigr\}
$$
Here $\nu=1,2,3$ (or $\a,\b,\g$) correspond to bands of xz, yz, xy nature;
$\e_{\nu,k}$ and $H_U$ are the zero band energies and the on-centre
interactions with parameters from \cite{21}. The interband interaction
$T_{\a\b}=\sum_{k,\s}4t_{\a\b}\sin{k_x}\sin{k_y}(\cg_{1 k\s}c_{2 k\s}
+h.c.)$ is small.  So in normal state with any spin structure one have three
almost independent bands with small  mixing of the $\a,\b$ bands at the
crossing
points of their FS's.  We neglect this mixing. For a sake of simplicity we
retain the notation $\a,\b$ for the unmixed bands of xz, yz nature. Mutual
influence of one band to other are provided by common chemical potential and
by mean fields created by electrons of all bands. These fields are
spin-dependent due to the on-centre exchange interaction.  The interaction
$V_{<nm>}$ of neighbour sites of type (1) is included in order to model
the possible singlet and triplet pairing in system.

In MF approximation the energy averaged over an arbitrary BCS-like state is
an explicit function
\be
{\ol H}= {\ol H}^N (y_i) + {\ol H}^{SC}(w_j, \t_j)
\label{3}
\ee
depending on the normal ($y_i$) and anomalous ($w_i, \t_j$) one-electron
averages named as OP's.  Among the normal OP's $\{y_i\}$ there
are the on-centre ($l=0$) and bond ($l=e_x,e_y$) densities
$r^{\nu}(l)=<{1\over 2}\sum_{\s} \cg_{n,\nu ,\s}c_{n+l,\nu,\s}>$ in each band
$\nu$, the mean kinetic energies
$T^{\nu}(l)=<{1\over N}\sum_{k\s} \e_k\cg_{k\s}c_{k\s}>$
and the local ($l=0$) or bond ($l=e_x,e_y$) spin densities ${\bf d}^{\nu}(l)$.
The local spin densities
\be
d^{\nu}(0)=[d^{\nu}(0)]^*=< e^{iQn}\cg_{n\nu\ur}c_{n\nu\dr}>
\label{4}
\ee
determine the spiral spin structure with a spiral vector $Q=2\pi({1\over 3},
{1\over 3})$
\be
<{\bf S}_{n\s}>={\bf e}_x \cos{Qn}+{\bf e}_y\sin{Qn}
\label{5}
\ee
Previous calculations \cite{16} show that the energy of normal spiral
state with such Q is lower than the energies of the para-, ferro- and
antiferromagnetic MF solutions. Such structure removes the nesting instability
of the $\a,\b$ bands and the exchange fields induce the same spiral spin
structure in the
$\g$ band also. The MF solution gives the collinear contributions to the
on-centre local spin from each band.

Since the FS's are different for all three bands, we should consider only the
intraband anomalous averages determined by the large phase volume around the
whole FS. A formation of pairs $(\cg_{k\nu\ur}\cg_{-k\nu'\dr})$ from
different bands $\nu\neq\nu'$ would be effective only at small
regions of $k$ near the crossing points of their FS's. Contrary to such
arguments recently\cite{14} the interband pairing pairing mechanism
has been proposed for $Sr_2RuO_4$.  However, the comparative estimations of
corresponding pairing susceptibility contradicts the possibility of such
pairing. So we retain the intra-band SC OP's only and neglect the
interband pair scattering. Then the expression for ${\ol H}^{SC}$ is
\be
\begin{array}{ll}
{1\over N}{\ol H}^{SC}=&U|w_{\nu}(0)|^2+ \sum_{\nu,~l=e_x,e_y}\Bigl\{
\k^s_{\nu}(l)|w_{\nu}(l)|^2\\& +
\k^t_{\nu}(l)\sum_{\mu=0,\pm 1}{1\over {1+|\mu |}}|\t_{\mu,\nu}(l)|^2
\Bigr\}\\
\end{array}
\label{6}
\ee
Here the quantities $w(l),\t(l)$ are the singlet and triplet superconducting
order parameters (SC OP's) on one centre ($l=0$) or on the neighbour centres
($l=e_x$ or $e_y$). They are determined as
\be
w_{\nu}(l)={1\over {2N}}\sum_{n\s}{\s\over {|\s|}}<\cg_{\nu, n,\s}
\cg_{\nu, n+l,-\s}>
\label{7}
\ee
\be
\begin{array}{ll}
\t_{\nu\mu}(l)=&{1\over {2N}}\sum_{n\s}e^{i\mu Q(n+l/2)}(\s_{\mu}\s_y)_{ss'}\\
&<\cg_{\nu, n,s}\cg_{\nu, n+l,s'}>\\
\end{array}
\label{8}
\ee
Here matrices $\s_{\mu}$ are $\s_z$ or $\mp(\s_x\pm i\s_y)$ for
$\mu=0, \pm 1$, and $\s_{x,y,z}$ are the Pauli matrices. The  phases $\phi
(n,l)=\mu Q(n+l/2)$ for $\mu=0,\pm 1$ in definition of the triplet OP's (8)
provide that the contribution from each bond $<n,n+l>$ are independent on $n$
in the same manner as the cyclic spin components $<S_{n\mu}e^{-i\mu Qn}>$
for  $\mu=\pm 1$ are independent on $n$ for state with the
spiral spin structure. These phases are connected with existence of
the spin currents in the spiral state. As it
will be shown later the coupled triplet pairs $(\ur\ur)$ or
$(\dr\dr)$ with the spins $\mu=1$ or $-1$ occur to be a moving pairs carrying
large total momenta $\pm Q$ correspondingly.  The neighbour-centre pairing
constants $\k^{s,t}$  in (6) are connected with the constants $V(l)$, $J(l)$
for the supposed interaction of type(1).

Having in mind an essential independence of $\a,\b,\g$ bands, consider
the possible SC orders in each band separately in order to
study a symmetry of SC order compatible with the spiral spin
order. Besides, we confirm  the previous conclusion \cite{5} that the
most probable "active" band for SC instability is the $\g$ band. In this line
we neglect the
interband Cooper pair scattering and take into account only the intra-band
constants $\k^s$, $\k^t$ in (7). To reduce the number of the SC OP's we use
the arguments typical for all strongly correlated systems. At any interaction
$V_{<nm>}$ in (1) a large on-centre interaction $U>0$ suppresses the singlet
s-wave pairing $w(0)$ in $\g$ band according to Eq.(6). So we retain only
the $d_{x^2-y^2}$-wave OP among the singlet OP's and take
$w_{\g}(0)=0$,  $w_\g (e_x)=-w_\g (e_y)$.
This provides the pair function which is orthogonal to the forbidden s-wave
pair function. For quasi 1D $\a,\b$ bands (here $\a,\b$ refer to xz, yx bands
rather than their combinations) the same on-centre interaction suppresses all
singlet pairings and we put $w_{\a(\b)}(0)=w_{\a}(e_x)=w_{\b}(e_y)=0$ since
both combinations $ w_{\a}(e_x)\pm w_{\a}(e_y)$ are non-orthogonal to the
on-centre pair function $w(0)$ for the quasi- 1D bands with a broken
tetragonal symmetry.

The neighbour site interaction of type (1) can induce the SC order only if
some of the constants  $\k_{\nu}^s$, $\k_{\nu}^t$ in  (6) are negative. Such
"attraction" should be thought rather as that of a  kinematic or correlational
nature or due to the hybrid character of "site orbits" composed from the d-
and $p_{\pi}$ orbits of ruthenium and oxygen. For this reason we suppose that
corresponding parameters $\k^s_{\nu}(l)$, $\k^t_{\nu}(l)$, or equivivalently
$V_{\nu}(l)$, $J_{\nu}(l)$ in model interaction (1), have non-zero values only
for those bonds $<nm> ,~m=n+l,$ for which the hoping $t^{\nu}_{nm}$ is
non-zero. This assumption is inspired by the expression for the exchange
interaction $J_{nm}\sim 4t^2_{nm}/U$ in t-J model. So
among all possible constants we set the non-zero values only for
the following singlet (s) and triplet (t) pairing constants
\be
\k^{s(t)}_{\a} (e_x)=\k^{s(t)}_{\b} (e_y)
=\k^{s(t)}_{\g} (e_x)=\k^{s(t)}_{\g} (e_y)=\k^{s(t)}
\label{9}
\ee
In the same analogy with t-J model one would expect the signs
$\k^s=-\k^t<0$.  In connection  with an expected triplet SC order in
$Sr_2RuO_4$ \cite{1} we especially extend our calculations on two limiting
cases with both signs of the triplet constant
\be
I.~~\k^s=-\k^t<0;~~~~~~~~~~~~II.~~\k^s=\k^t<0
\label{10}
\ee
First case corresponds to an attraction of particles in the singlet channel,
but repulsion in the triplet one. The second case corresponds to attraction in
both channels.

The MF procedure is the standard one. From explicit function ${\ol H}(z_i)$
of mean energy depending on the one-particle  OP's $z_j=\{y_i , w_i , \t_i
\}_j$ one obtains the linearised Hamiltonian
\be
H_{Lin}-\mu N={ {\p ( {\ol H}-\mu{\ol N})} \over {\p z_i}}
({\hat z}_i -z_i) +  {\ol H}(z_i)-\mu{\ol N}
\label{11}
\ee
where ${\hat z}_i$ are the one-particle operators corresponding to respective
OP's $z_i$.  The MF BCS-like state is the eigen state of $H_{Lin}$ and in turn
this state determines the values of OP's $z_i$. Thus the selfconsistent
solution is obtained.

For the state with the spiral spin structure the most convenient  basis set
for representation of (11) is a basis of the following Fermi operators
\be
\bg_{i,\nu, k}=\Bigr\{\cg_{\nu, k, \ur},~\cg_{\nu, k+Q, \dr},
~c_{\nu, -(k+Q), \ur},~c_{\nu, -k, \dr}\Bigl\}_i,
\label{12}
\ee
for each band $\nu$. Here $i=1,\dots, 4$ and the momentum $k$ runs inside the region $F$ which
is   a half of whole momenta space and is limited  by conditions
\be
k\in F:~~~~~(k+Q/2)Q <0
\label{13}
\ee
For vector Q with $Q_x=Q_y=2\pi/3$ the Eq.(13)  means that the components
$k_x,~~k_y$ vary in limits
$$
-\pi-Q_x/2<k_{x,y}<\pi-Q_x/2;~~~~~k_x+k_y+Q_x<0
$$
The linearised Hamiltonian  $H_{Lin}$ is determined by the independent
contributions in a basis subset (12)
\be
H_{Lin}=\sum_{k\in F}{\hat h}_k^{\nu}
;~~~~~~~{\hat h}_k^{\nu}=h_{ij}^{\nu} \bg_{i,\nu,k}b_{j\nu,k}
\label{14}
\ee
Here the fourth rank Hermitian  matrix $h_{ij}^{\nu}$ is determined by
elements (the omitted band index $\nu$ is implied here)
\be
\begin{array}{ll}
h_{11}=\e_{\nu} (k_1)-\mu;&h_{22}=\e_{\nu} (k_2)-\mu;\\
h_{12}=-h_{24}=-{1\over 2}\p{\ol H}/\p d_{\nu};~&
h_{14}=A(k_1)+B_{0}(k_1);\\
h_{23}=-A(k_2)+B_{0}(k_2);&
h_{13(2,4)}=B_{\pm 1}({\ol k});\\
h_{33}=-h_{22};&h_{44}=-h_{11}\\
\end{array}
\label{15}
\ee
where
$$
~~~k_1=k, ~~~ k_2=k+Q, ~~~~{\ol k}=(k_1+k_2)/2
$$
and the functions $A,~B_{\mu}$ are
\be
\begin{array}{ll}
A(k)&=\sum_{l=e_x,e_y}\k_{\nu}^s (l)w_{\nu}(l)\cos{kl};\\
B_{\mu} (k)&=\sum_{l=e_x,e_y}\k_{\nu}^t (l)\t_{\mu,\nu}(l)\sin{kl}\\
\end{array}
\label{16}
\ee
The d-wave symmetry of singlet OP in $\g$ band requires the antisymmetry of
$A^{\g}(k_x,k_y)=-A^{\g}(k_y,k_x)$ at $k_x\lrr k_y$. The MF
solution reveals also the same values of $\t_{\mu}$ for $\mu=\pm 1$. So
actually only three non-zero real order parameters
determine all SC OP's in $\g$ band. They are
$$z_i=(w_d,~\t_0,~\t_1)_i,~~i=1,2,3$$
\be
\begin{array}{ll}
w_d=&{1\over 2}[w(e_x)-w(e_y)],~\t_0={1\over 2}[\t_0(e_x)-\t_0(e_y)]\\
&\t_1={1\over 4}\sum_{\mu=\pm 1}[\t_{\mu}(e_x)-\t_{\mu}(e_y)]
\end{array}
\label{17}
\ee
The same symmetry with respect to $x\lrr y$ should be applied to  the
solutions in $\a,\b$ bands, but with simultaneous replacement $\a\lrr \b$
($xz\lrr yz$). The same values of $\t_{\mu}$ for both projections $\mu=\pm 1$
are revealed in MF solution in $\a,~\b$ bands also. So, for $\a,\b$ bands we
retain the next triplet OP's corresponding to the non-zero triplet constants
in (9):

\be
\t_{\mu,\a}(e_x)=-\t_{\mu,\b}(e_y);~~~~  |\mu|=0, ~1
\label{18}
\ee
It was verified that if initial values of OP's satisfies Eqs.(17,18), then
the subsequent iterations of selfconsisting MF procedure conserve the same
symmetry.

One more simplification has been used. Really the interaction of type
(1) gives the contribution to both parts ${\ol H}_N (y_i)$ and ${\ol H}_{SC}$
of the mean energy (3). We may consider that the first contribution depending
on the normal state charge and spin densities is already taken into account
in renormalised band energies $\e_{\nu}(k)$ fitted earlier \cite{21} to those
obtained from
the observed magnetic quantum oscillations. Thus we retain in ${\ol H}$ only
the part of $<V>$ which depends on anomalous averages. The definitions (8)
of the triplet OP's allows to deal with the real solutions. They have a
certain  symmetry relative to reflection in the diagonal
plane $(z,x=y)$ containing the spirality vector $Q$ with simultaneous exchange
of  bands $\a \lrr \b$ and relative to reflection in plane $(z,x=-y)$
($Q\to -Q$) together with $\s\to -\s$.

The BCS-like spiral state under a search is determined by filling the
one-particle eigen states $\h^{\dg}_{\l \nu k}$ corresponding the energy
levels $E_{\l}(k)$
\be
\h^{\dg}_{\l \nu k}=\bg_{ik}S_{i\l}(k);~~
h_{ij}(k)S_{j\l}(k)=S_{i\l}(k)E_{\l}(k)
\label{19}
\ee
The matrices $S_{i\l}(k)$  of  eigen-vectors and the Fermi occupancies
$f(E_{\l\nu k})$ of levels determine the normal and anomalous OP's (7,8),
thus  closing the selfconsisting MF procedure.
\section{The results.}
Since a full MF solution with
SC order is easily obtained at large pairing constants, we first study the
models with large $k^{s(t)}$. Then we present the phase curves $T_c(k^s)$ and
relative values of SC OP's for realistic models with small $k^s$.

The results are obtained for two cases (10) of pairing constants
corresponding to attraction in a singlet channel only (case I) or in both
the singlet and triplet channels (case II).
In first case the $\a,~~\b$ bands do not display any own SC order. The reason
is that both the singlet "d-wave", as well as "s-wave" OP are suppressed by
on-centre interaction in bands with nonequivalent hopping in $x$ and $y$
direction:  $t_x^{\a}\gg t_y^{\a}$ or  $t_y^{\b}\gg t_x^{\b}$  (see parameters
of three-band model in \cite{21}). In difference, in $\g$ band of system in
spiral spin configuration the mixed SC order arises. The d-wave singlet order
is accompanied by the triplet pair formation
even at $\k^t_{\g}>0$. Fig.1 shows the temperature dependence of the singlet
and  triplet OP's (17) of $\g$ band for $\k^s=-\k^t=-0.6 eV$. The value of
triplet OP's occur to be $\t_{+1,\g}=\t_{-1,\g}\gg \t_{0,\g}$. Taking into
account the definition of the triplet OP's in (17,18) and their momentum
representation one concludes that the coupled triplet pairs of particles of
$\g$ band arise mainly in form $(\ur\ur)$ or $(\dr\dr)$ and these pairs are
the moving coupled pairs carrying a large total momenta $-Q$ or $Q$
correspondingly. This differs the triplet pairing in   spiral state from that
in isotropic Fermi liquid where only Cooper- like pairs $(k\ur,-k\ur)$ or
$(k\dr,-k\dr)$ with zero momenta can exist. Moreover, unlike the isotropic
model the triplet pairs arise even at positive value of triplet constant
$\k^t>0$ corresponding to repulsion in triplet channel.

Note that the emergence of coupled pairs with large total momentum $2k_F$
equal to a nesting vector was grounded in new theory of HTSC \cite{23}.
In \cite{23} such pairs are the singlet pairs associated with the stripe
structure. In case of the spiral state such moving pairs are the triplet ones.
Note, that the
photoemission technique with the circularly polarised photons recently
applied to cuprates \cite{22} could distinguish the supposed spiral structure
which breaks a time reversal symmetry even in the normal state. In particular
the sharp or smoothed (gaped) Fermi surfaces are predicted for electrons
$(k\s)$ with momentum $k$ from region $kQ<0$ (or $kQ>0$) and with polarisation
$\s=\ur$ (or $\dr$) correspondingly. Really the spin order have a local
finite-ranged character. Any disorder or any domain structure must suppress
the above mention polarisation effects.

Thus an attraction only in a singlet channel leads to both the singlet
and triplet pair formation in state with the spiral spin structure caused by
nesting of $\a$ and $\b$ bands. Fig.1 shows also the heat capacity of system.
A finite limit $C(T)/T$ at $T\to 0$ is due to contributions from $\a$, $\b$
bands for which the normal state is retaining if the interband Cooper pair
scattering is neglected.

Consider now the model of second type with  "attraction" in both channels:
$\k^s=\k^t<0$ in (10) .  At $\k^s>0.65eV$  all
three bands display their own SC order. At equal values of
the pairing constants (9) in each bonds with large hopping the  SC
order in $\g$ band has much pronounced character than that in $\a$ and $\b$
bands.  The ensemble of coupled pairs in $\g$ band consists mainly of the
d-wave singlet Cooper pairs $\{k\ur,-k\dr\}^s$ and the moving triplet pairs
$\{k\ur,-(k+Q)\ur\}$ or $\{k\dr,-(k-Q)\dr\}$ with total momentum $-Q$ or $Q$
correspondingly. In $\a,~\b$ bands the triplet order is providing mainly by a
Cooper-like triplet pairs $\{k\ur,-k\dr\}^t$ with a zero total
momentum.  They correspond to $\t_0^{\a}(e_x)=-\t_0^{\b}(e_y)\neq 0$. Fig.~2
illustrates the temperature dependence of
the SC OP's in $\a,~~\b,~~\g$ bands at large trial
constants $\k^s=\k^t=-0.8$eV. Note, that a relation $|\t_0(l)|>|\t_{\pm
1}(l)|$  of triplet components for $l=e_{x(y)}$ in $\a(\b)$ band differs from
a relation $|\t_0(l)|\ll |\t_{\pm 1}(l)|$ for $\g$ band. This is connected
with  a difference in the Fermi surfaces and in the angular dependence of
pairing potentials. Thus, our calculations confirm the conclusion \cite{5}
that the active band in system is the $\g$ band.  The independent SC
 transitions in different bands are accompanied by two steps in the
temperature dependence of a heat capacity. It is seen that $C(T)/T\to 0$ at
$T\to 0$ as it must be for the totally SC state. But since the scales of the
SC OP's are different for $\g$ and $\a,\b$ bands a decrease of $C(T)/T$  at
$T\to 0$ occurs at small T. Real situation should greatly depend on the
inter-band pair scattering neglected here.

The pairing potential in active $\g$ band with the SC OP's (17) may be
presented in form
\be
\begin{array}{ll}
&H^{SC}_{Lin}=\sum_{k\in G}\{[A(k)+B_0(k)]\cg_{k\ur}\cg_{-k\dr}
+B_1(k)\times\\
&[\cg_{k-Q/2,\ur}\cg_{-k-Q/2,\ur}+\cg_{k+Q/2,\dr}\cg_{-k+Q/2,\dr}]+h.c.\}\\
\end{array}
\label{20}
\ee
Here $k$ runs the whole phase space $G$ (in difference from representation
(14), where $k$ runs only half of $G$) and the omitted index $\nu=\g$ is
implied. Functions $A(k),~B_{\mu}(k)$ are defined by Eqs.(16,17) and have
a form
\be
\begin{array}{ll}
A(k)=&\k^sw_d(\cos{k_x}-\cos{k_y});\\
B_{\mu}(k)=&\k^t\t_{\mu}(\sin{k_x}-sin{k_y})\\
\end{array}
\label{21}
\ee
They obey the symmetry $A(k_x,k_y)=-A(k_y,k_x)$ and
$B(k_x,k_y)=-B(k_y,k_x)$ .  Therefore in any cases
the diagonal line $k_x=k_y$ along the vector $Q$ is the node line of the
gap function in our solution. Thus, in spiral state both the d-wave singlet
and p-wave triplet SC orders are compatible. In case of isotropic normal
state without the spiral spin structure ($d_{\nu}=0,~Q=0$) the pairing
potential (21) would be a superposition of contributions corresponding to
different representations of tetragonal point group classified in \cite{5a}.
In notations \cite{5a} this superposition may schematically be presented as
$\{\t_1(A_{1u}-B_{1u}-A_{2u}+B_{2u})+\t_0 (E_{ux}-E_{uy})\}$.
In difference from paramagnet state in spiral state the coupled pairs
$(\ur\ur)$,
$(\dr\dr)$ are the pairs with large total momentum $\mp Q$. The spin currents
$j_{\ur\ur}=-j_{\dr\dr}$  due to the pair motions are of the same signs as
the spin currents $j_{\ur}=-j_{\dr}$ in the normal state with the spiral
structure. Remind, that the spins are projected on the axis $z$ perpendicular
to the spin rotation plane of the spiral ground state structure (5).

The gap function in the $\g$ band corresponding the paring potential (21)
can be defined as a real matrix element
\be
G(k)=< \eta_{-k} | H^{SC}|\eta^{\dg}_k>
\label{22}
\ee
between the electron and hole quasiparticles $\eta^\dg,\eta$
of the upper Hubbard band of the normal spiral state.

Fig.~3 present the map of levels of the gap function $G(k)$ of $\g$ band for
case $\k^s=-k^t<0$. The gap function is antisymmetric relative to
an exchange $k_x\to k_y$ but does not possess the inversion symmetry. However,
an imagine  photoemission experiment for the $\g$ band could measure two
different gaps $|G(k)|$ and $|G(-k)|$ at each k for different polarisations
(up and down) of the emitted electrons. Note, that along the known FS of $\g$
band the function $G(k)$ is close to the d-wave function
$\sim\cos{k_x}-\cos{k_y}$. This is consistent with the observed 4-fold
anysotropy of heat conductance in the in-plane magnetic field \cite{9,10}.

Up to now the results refer to models with too large pairing constants and
$T_c$. For more realistic models with small $T_c$ and the pairing constants
of both types (10) we carry out the calculations of the phase curves
$T_c(\k^s)$. Dependence $T_c(\k^s)$ of the SC transition temperature in $\g$
band is determined by equation following from linear expansion of r.h.s. of
(7,8) over $w_{\g},\t_{\g \mu}$
\be
Det\parallel \d_{ij}-R_{ij}\k_j\parallel=0
\label{23}
\ee
Here $i,j$ numerate the SC OP's $\{z_i\}$ and a matrix $R_{ij}$ is
\be
R_{ij}={1\over N}\sum_k \sum_{\l,\l'} M^i_{\l,\l'}M^j_{\l,\l'}
{{f(-E_\l)-f(E_{\l'})}\over{E_\l+E_{\l'}}}
\label{24}
\ee
where $E_\l,f(E_\l)$ are the normal state energies and the Fermi occupancies.
Matrices $M^i$, $i=1,2,3$, corresponding to SC OP's (17) are given
in Appendix and constants $\k_j$ in (23) are $\k_j=\{\k^s,\k^t,\k^t\}_j$.
The indices $\l,\l'=1,2$ numerates the normal state levels
in the upper and lower subband
of $\g$ band. At the transition points $T=T_c$, when Eq.(23) is satisfied,
the corresponding homogeneous equations
\be
(\d_{ij}-R_{ij}\k_j){\tilde z}_j=0; ~~~
{\tilde z}_j=z^\g_j/\sqrt{z_1^2+z_2^2+z_3^2}
\label{25}
\ee
determine the relative normalised values ${\tilde z}_j$ of the SC OP's.
Fig.~4 presents the phase curves $T_c(\k^s)$ and relative values of OP's
${\tilde z}_j$ at $T\to T_c$ as function of $\k^s$ for two signs of triplet
pairing constant $\k^t$ in (10). Points mark the values ${\tilde z}_j$ for
$T\sim 0.6T_c$ obtained from full MF calculations for the above studied
models. These values are in consistency with those obtained from Eqs.(25).
The models with a realistic small $T_c\sim 10^{-4}$eV display
the same symmetry properties of SC state as the models with large $\k^s$ and
$T_c$. The transition temperature $T_c=1.5K$ observed in $Sr_2RuO_4$
corresponds to values $\k^s=0.145~eV$ or $\k^s=0.12~eV$ for each types (10)
of models.

Many important problems concerning $Sr_2RuO_4$ remain out of the present
study, for example a proximity effect, i.e. a mutual influence of different
bands via the interband pair scattering \cite{5}. This effect should be
included in order to describe the observed common SC transition in all three
bands. Its relation to the thermodynamics and magnetic properties have been
discussed intensively \cite{5,14,16}.

\section{Conclusions}
The above model treatment leads to following conclusions.
The SC pairing in $\g$ band can coexist with a spiral spin order caused by
the nesting in $\a,\b$ bands of $Sr_2RuO_4$. The most probable "active" band
relative to the SC instability is the $\g$ band. The mixed d-wave singlet and
p-wave triplet SC order arises from the pairing interaction of adjacent
sites on background of the normal state with the spiral spin structure
described by the nesting vector $Q=2\pi (1/3,1/3)$. For both types of pairing
constants - with attraction in the singlet and triplet channels or only in
the singlet one - the main coupled pairs in system are the singlet d-wave
pairs $(k,-k)_{\ur\dr}^s$ and the moving triplet
pairs $(k-Q/2,-k-Q/2)_{\ur\ur}^s$ , $(k+Q/2,-k+Q/2)_{\dr\dr}^s$ with large
total momenta $\mp Q$ and with the spin projections $\mu=\pm 1$ on an axis
normal to the spin-rotation plane of the spiral structure.
The preferential d-wave pairing in $\g$ band is consistent with the
observed 4-fold anysotropy of heat conductance in the in-plane magnetic field
\cite{9,10}. Only triplet SC order may be transfered from $\g$ to
$\a,~\b$ bands via interband pair scattering.
The studied spiral normal and
SC states are both the states with a broken time reversal symmetry. The
effects of symmetry breaking might be detected in photoemission with
circularly polarised light.
The important questions remain unsolved: Can the mixed singlet-triplet
pairing induced by attraction in singlet channel provide an independency of
the Knight shift on temperature at $T<T_c$ ?  Can the other periodic
structures (besides the spiral one) exist in the normal state of $Sr_2RuO_4$
and what is a relation between the triplet and singlet SC pairing in these
structures?

Work is supported by Russian Fund of Fundamental Research (Projects
No. 00-03-32981 and No. 00-15-97334). Authors are grateful to  V.Ya Krivnov
for useful discussions and P.Fulde for possibility to work in Max Planck
Institute for Physics of Complex Systems, Dresden.  We thank also
W.Werner for presenting preprints of his work before its publication.  %
\section{Appendix}
The matrices $M^i$ in (24) are given by
\be
M^i_{\l,\l'}(k)=\Biggl[
\Biggl( \begin{array}{cc}c&s\\s&-c\\ \end{array}\Biggr)
{\ol M}^i
\Biggl( \begin{array}{cc}s&c\\c&-s\\ \end{array}\Biggr)
\Biggr]_{\l\l'}
\label{26}
\ee
$$
\begin{array}{ll}
&{\ol M}^1=
\Biggl( \begin{array}{cc} 0&c_d(k_1)\\-c_d(k_2)&0\\ \end{array}\Biggr),~~
{\ol M}^2=
\Biggl( \begin{array}{cc}0&s_p(k_1)\\s_p(k_2)&0\\ \end{array}\Biggr),~\\
&~~~~~~~~~~~~~{\ol M}^3= s_p({\ol k})
\Biggl( \begin{array}{cc}1&0\\0&1\\ \end{array}
\Biggr)\\
\end{array}
$$
Here $s=\sin{\f}$, $c=\cos{\f}$ and  $\f(k)$ for band $\nu$ is determined
by equation $\tan{(2\f)}=-[{\p{\ol H}/\p d_{\nu}][\e(k_1}-\e(k_2)]^{-1}$.
The other functions are $c_d(k)=(\cos{k_x}-\cos{k_y})/2$,
$s_p(k)=(\sin{k_x}-\sin{k_y})/2$ and $k_1=k$, $k_2=k+Q$, ${\ol k}=k+Q/2$.
%
%


\newpage
\vspace {0.1in}
{\bf Captions to Figures}
\vspace {0.05in}

\begin{figure}
\caption{
The total heat capacity devided by $T$ - $C(T)/T$ (in $[eV^{-1}]$) and  the
SC OP's in $\g$ band as functions of temperature for model with independent
bands and large pairing constants $\k^s=-\k^t=-0.6 eV$. Curves 1-3 refer to
$w_d$, $-\t_0$ and $-\t$ determined by Eq.(17).
}
\end{figure}
%
\begin{figure}
\caption{
The same as in Fig.~1 for model with $\k^s=\k^t=-0.8 eV$. Curves 1-3 refer to
the same OP's of $\g$ band as in Fig.~1, curves 4,5 refer to
$\t_0^{\a}(e_x)=-\t_0^{\b}(e_y)$ and $\t_{\pm 1}^{\a}(e_x)=-\t_{\pm
1}^{\b}(e_y)$.
}
\end{figure}
%
\begin{figure}
\caption{
The contour map of the gap function (23) for the same model as in Fig.~1
in a momentum space $|k_{x(y)}|<\pi$.
Solid (dashed) levels refer to positive (negative) values of gap function
$G(k_x,k_y)$.
}
\end{figure}
%
\begin{figure}
\caption{
Phase curves $T_c(\k^s)$ and the relative normalised values of OP's
(25,17) of $\g$ band as functions of the singlet pairing constant.  Solid
(dashed) curves refer to models with $\k^t=\pm \k^s$ correspondingly.  Curves
1,2,3 refer to ${\tilde z}_i$, $i=1,2,3$, obtained from solution
of homogeneous equations (27) ($T\to T_c$).  Points (circles and squares)
are the same values at $T=0.6T_c$ obtained from full MF calculations for the
the models which are presented in Fig.~1  and Fig.~2 correspondingly.
}
\end{figure}
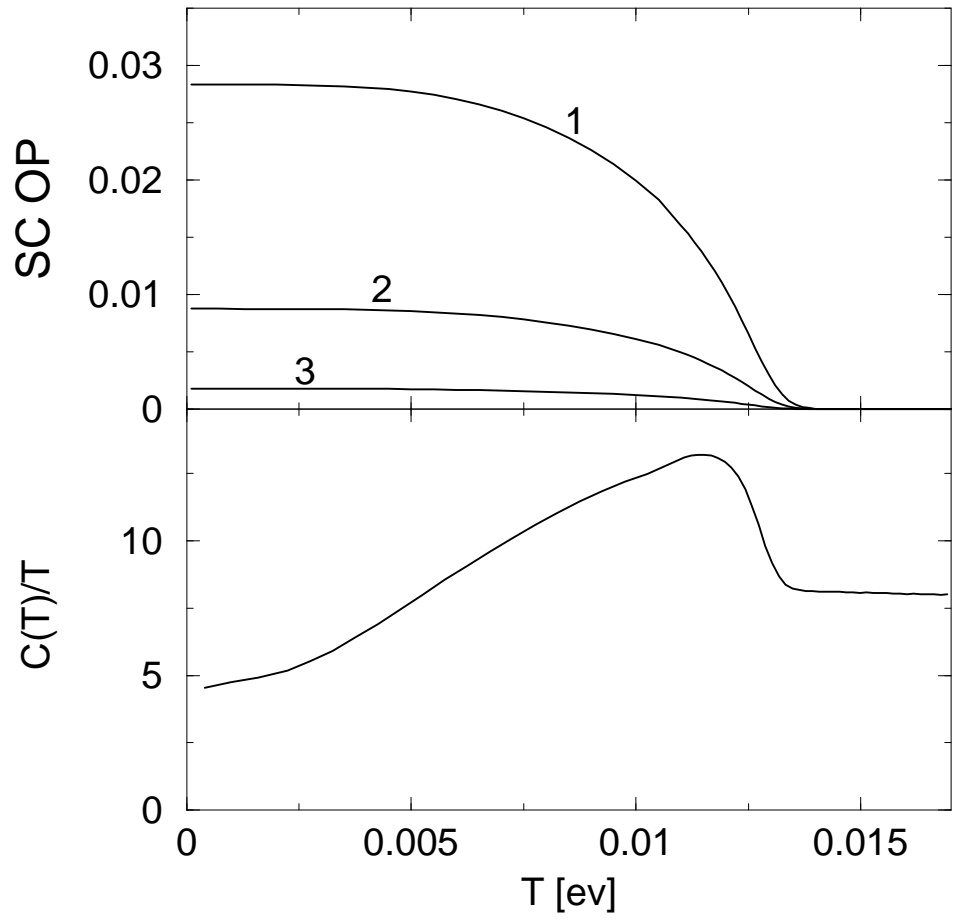
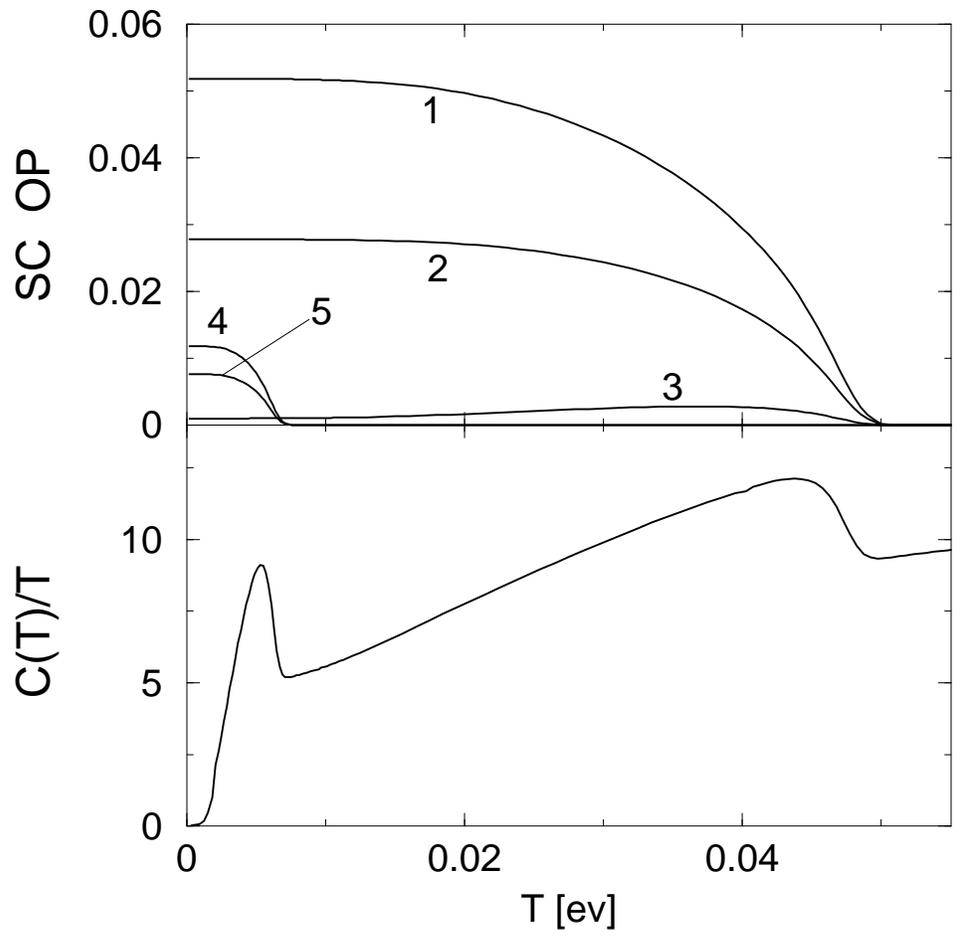
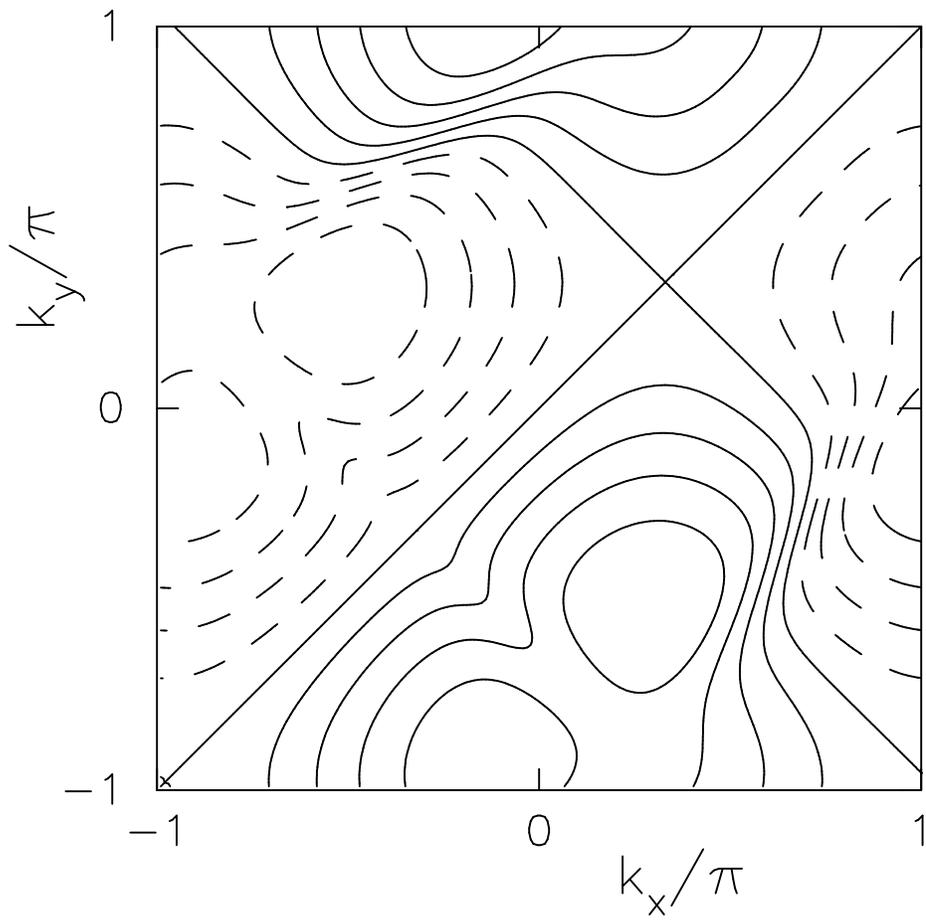
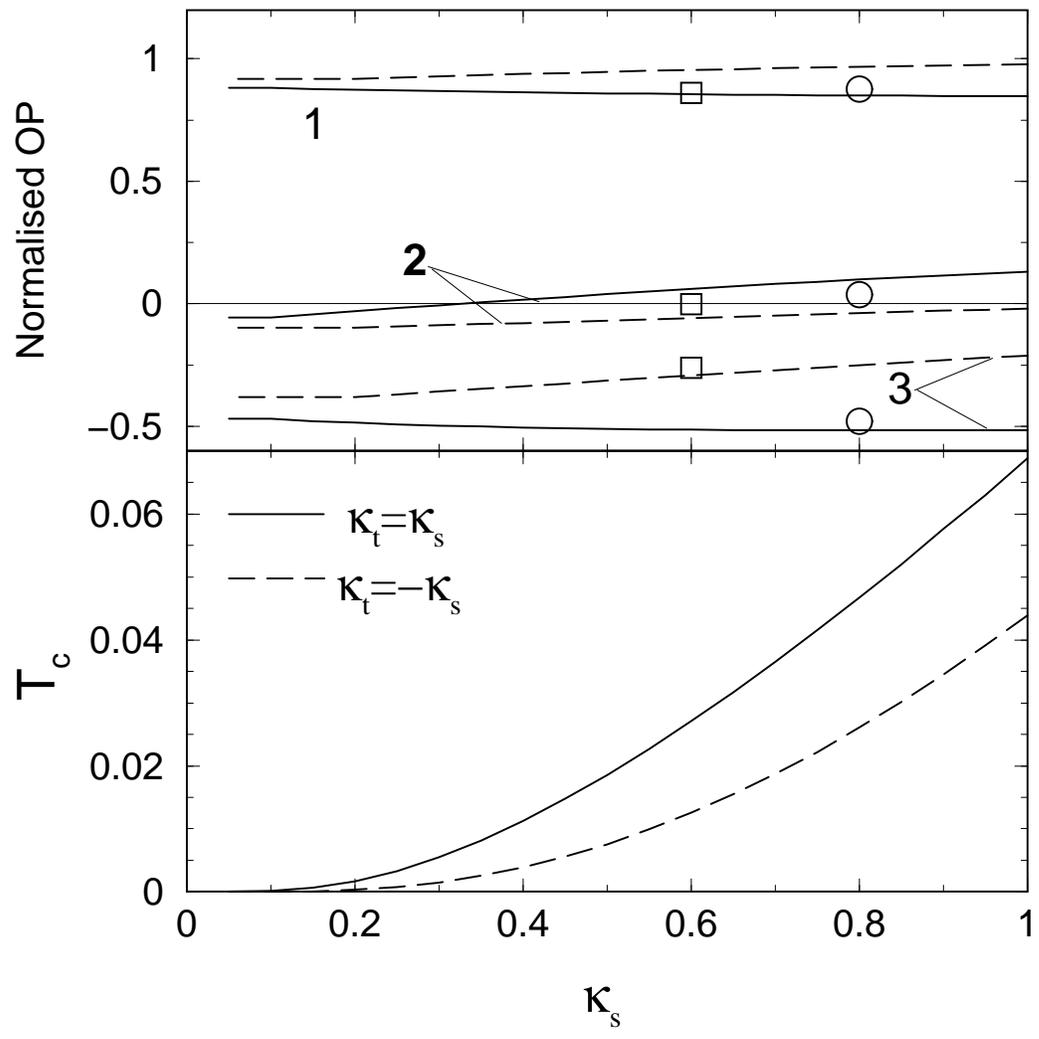
\end{document}